\documentstyle[preprint,eqsecnum,aps]{revtex}
\tightenlines
\begin{document}
\draft
\preprint{}
\title{On the Concept of Quantum State Reduction:\\
Inconsistency of the Orthodox View}
\author{Masanao Ozawa\thanks{Electronic address:
e43252a@nucc.cc.nagoya-u.ac.jp}}
\address{School of Informatics and Sciences,
Nagoya University, Nagoya 464-01, Japan}
\date{\today}
\maketitle
\begin{abstract}
The argument is re-examined that the program of deriving the rule of 
state reduction from the Schr\"{o}dinger equation holding for the 
object-apparatus composite system falls into a vicious circle
or an infinite regress called the von Neumann chain.
It is shown that this argument suffers from a serious physical 
inconsistency concerning the causality between the reading of the 
outcome and the state reduction.
A consistent argument which accomplishes the above program without 
falling into the circular argument is presented.
\end{abstract}
\narrowtext
% SECTION 1:
\section{Introduction}\label{se:1}
\setcounter{equation}{0}

Quantum mechanics includes a dualism 
concerning the principle of state change.
The Schr\"{o}dinger equation, on the one hand, governs 
the state change caused by time evolution.
The rule of state reduction, on the other hand, governs 
the state change caused by measurement.
The dualism is justified as long as 
the state change of only one system is concerned.
The Schr\"{o}dinger equation holds true only when the system is isolated, 
but every measurement accompanies 
the interaction with the measuring apparatus so that 
the rule of state reduction holds true only when the system is not isolated. 
The dualism is therefore justified \cite[p.~420]{vN55}.

Accepting that every measurement accompanies 
the interaction between the object and the apparatus at all, 
one can expect that the rule of state reduction can be derived from 
the Schr\"{o}dinger equation holding for the composite system of 
the object and the apparatus during the measurement.
A negative view, however, prevails against this 
program.
According to that view,
the Schr\"{o}dinger equation for the composite system transforms
the problem of a measurement on the object to the problem of an observation
on the apparatus, 
but in order to derive the rule of state reduction 
holding for the object one still needs the rule of state reduction 
applied to the composite system \cite[p.~329]{Wig63}.
This implies that the program of deriving the rule of state reduction
from the Schr\"{o}dinger equation holding for the object-apparatus
composite system falls
into a vicious circle sometimes called von Neumann's chain
\cite[Section 11.2]{Jam74}.

The purpose of this paper is to show that the above argument,
usually called the orthodox view of measurement theory,
includes a serious physical 
inconsistency and then to present a consistent argument 
which derives the rule of state reduction from
the Schr\"{o}dinger equation of the composite system
without falling into the vicious circle.

In this paper, we are confined to the state reduction caused
by a measurement of an observable with nondegenerate purely
discrete spectrum satisfying the repeatability hypothesis. 
Sections \ref{se:2}--\ref{se:7} review with elaboration the most basic part 
of measurement theory originated with von Neumann \cite{vN55}.
Section \ref{se:2} presents postulates for quantum mechanics and defines 
state reduction.  Section \ref{se:3} introduces the notion of nonselective
measurement and shows that a nonselective measurement causes a
state change in quantum mechanics whereas it is not the case in
classical mechanics.  Section \ref{se:4} concludes the existence of an
interaction between the measured object and 
the apparatus in every measurement. 
Section \ref{se:5} shows that the rule
of state reduction is equivalent to the repeatability hypothesis.
Section \ref{se:6} introduces the projection postulate as the rule of
state reduction in the case where the observable has degenerate
spectrum.  Section \ref{se:7} derives a necessary condition for a unitary 
operator to represent the measuring interaction.
The condition determines the form of the unitary operator 
representing the measuring interaction leading to state reduction.  
The problem is then formulated as whether the unitary operator 
of this form is sufficient for deriving the rule of state reduction.
Section \ref{se:8} reviews the orthodox view along with Wigner's 
argument \cite{Wig63} that claims that the unitary
operator does not lead to the rule of state reduction without appealing
to the rule of state reduction, the projection postulate, 
for the measurement of the pointer position.
Section \ref{se:9} shows that the orthodox view suffers from a serious 
physical 
inconsistency concerning the causality between the reading of the 
outcome and the state reduction.
Section \ref{se:10} presents a consistent argument which derives the rule of 
state reduction from the Schr\"{o}dinger equation holding for the 
object-apparatus composite system without appealing to the projection
postulate for the pointer-measurement.
Concluding remarks are provided in Section \ref{se:11}.

\section{Postulates for quantum mechanics}
\label{se:2}

In this paper, all state vectors are supposed to be normalized, 
and mixed states are represented by density operators, i.e., 
positive operators with unit trace.
Let $A$ be an observable with a nondegenerate purely discrete
spectrum.  Let $\phi_{1},\phi_{2},\ldots$ be a complete orthonormal
sequence of eigenvectors of $A$ and 
$a_{1},a_{2},\ldots$ the corresponding eigenvalues; by
assumption, all different from each other.

According to the standard formulation of quantum mechanics,
on the result of a measurement of the observable $A$ the following
postulates are posed:

(A1) {\em If the system is in the state $\psi$ at the time of measurement,
the eigenvalue $a_{n}$ is obtained as the outcome 
of measurement with the probability $|\langle\phi_{n}|\psi\rangle|^2$. }

(A2) {\em If the outcome of measurement is the eigenvalue $a_{n}$,
the system is left in the corresponding eigenstate $\phi_{n}$ at the
time just after measurement.}

The postulate (A1) is called the {\em statistical formula}, and
(A2) the {\em measurement axiom}.
The state change $\psi\mapsto\phi_{n}$ described by the measurement 
axiom is called the {\em state reduction}.

\section{Origin of measurement theory}
\label{se:3}

The state reduction can be thought to be caused by the following
two factors: the dynamical change of the system and the change of 
the observer's knowledge.
In order to separate these factors, let us suppose that we terminate
the procedure of measurement of the observable $A$ just before the
observer's reading of the outcome; this procedure is called a 
{\em nonselective measurement}.
It follows from postulates (A1) and (A2) that the nonselective measurement
leaves the system in the mixture of the states 
$|\phi_n\rangle\langle\phi_n|$ with the probability
$|\langle\phi_n|\psi\rangle|^2$ and therefore yields the following state 
change: 
\begin{equation}\label{eq:0}
|\psi\rangle\langle\psi| \mapsto 
\sum_{n} |\langle\phi_n|\psi\rangle|^2\,|\phi_n\rangle\langle\phi_n|.
\end{equation}
Since a nonselective measurement does not change the observer's
knowledge, this state change is considered to be caused entirely 
by the dynamical factor.  

Even in classical mechanics, when the outcome of measurement
is obtained, 
the information on the outcome changes the observer's knowledge
and the probabilistic description of the state of the
system changes according to the Bayes principle, which is
formulated as follows:
Let $X,Y$ be two (discrete) random variables.
Suppose that we know the joint probability distribution
$\Pr\{X=x,Y=y\}$.
Then, the prior distribution of $X$ is defined as the 
marginal distribution of $X$, i.e., 
\begin{equation}
\Pr\{X=x\}=\sum_{y}\Pr\{X=x,Y=y\}.
\end{equation}
If one measures $Y$, the {\em information} ``$Y=y$'' changes the probability 
distribution of $X$ for any outcome $y$.
The posterior distribution of $X$ is defined as the 
conditional probability distribution of $X$ given $Y=y$, i.e., 
\begin{equation}
\Pr\{X=x|Y=y\}=\frac{\Pr\{X=x,Y=y\}}{\sum_{x}\Pr\{X=x,Y=y\}}.
\end{equation}
This principle of changing the probability distribution from the
prior distribution to the posterior distribution is called
the {\em Bayes principle}.

Nonetheless, a nonselective measurement in classical mechanics
causes no change in the system.
Therefore, it is a characteristic feature of quantum mechanics
that a nonselective measurement changes the system dynamically,
and it is the origin of von Neumann's measurement theory to explain 
this change.

\section{Existence of measuring interaction}
\label{se:4}

In quantum mechanics the state of an isolated
system changes dynamically according to the Schr\"{o}dinger equation,
but this state change does not change the entropy of 
the system. 
On the other hand, the state change (\ref{eq:0}) caused by the nonselective
measurement increases the entropy of the measured system, and hence 
this process of state
change cannot be described by the Schr\"{o}dinger equation of the
measured system \cite[p.~388]{vN55}.
It follows that this dynamical change of state must be caused by 
the interaction between the measured object and the measuring apparatus,
a system external to the measured object including every system interacting
with the measured object.
Thus, from the basic postulates of quantum mechanics
we have derived the existence of measuring interaction, which is
neglected in classical mechanics.
Since our discussion concerns only nonselective 
measurements, we do not need to mention the function of consciousness,
although the conventional argument mentions 
the psycho-physical parallelism \cite[pp.~418--420]{vN55}.

\section{Repeatability hypothesis and state reduction}
\label{se:5}

In the measurement axiom (A2),
state reduction is described as a change of the state of the object.
In order to consider state reduction together with the interaction
between the object and the apparatus,
it is desirable to describe it independently of particular 
descriptions of states of systems.
As one of such descriptions,
von Neumann showed that (A2) is equivalent to the following
{\em repeatability hypothesis} \cite[pp.~213--218, 335]{vN55}:

(M) {\em If a physical quantity is measured twice in succession in
a system, then we get the same value each time.}

In fact, according to (M) the state of the object just after the 
first measurement is the eigenstate corresponding to the outcome,
and in the nondegenerate case it is determined uniquely so that
we have (A2).  It is obvious that (M) follows from (A2).

\section{Projection postulate}
\label{se:6}

In this paper, we are devoted to measurements of observables with
nondegenerate discrete spectrum.  
In the conventional discussion explaining
state reduction, however, we need to consider a measurement on
the object-apparatus
composite system and we need the statistical formula
and the measurement axiom for observables with degenerate spectrum.
The statistical formula for the discrete observable 
$X=\sum_{n}a_{n}E_{n}$ to be measured in the state (density operator) 
$\rho$ is given as follows:

(B1) {\em The eigenvalue $a_{n}$ is obtained as the outcome with
the probability $\mbox{\rm Tr}[E_{n}\rho]$.}

In the degenerate case, the eigenstate corresponding to an eigenvalue
is not determined uniquely, and hence (A2) is ambiguous.
Moreover, the repeatability hypothesis (M) implies that the object
is left in an eigenstate corresponding to the outcome, but (M) does
not determine the eigenstate in which the object is left.
For determining this eigenstate,
von Neumann posed no special principle \cite[p.~348]{vN55},
but later L\"uders proposed a principle which has been widely accepted
\cite{Lud51}.
According to his principle, 
if the measurement of the discrete observable $X$ with spectral 
decomposition $X=\sum_{n}a_{n}E_{n}$ is carried out in the state 
(density operator) $\rho$ and leads to the outcome $a_{n}$,
then:

(B2) {\em At the time just after measurement, the object 
is left in the state $E_{n}\rho E_{n}/\mbox{\rm Tr}[E_{n}\rho]$.}

This principle is called the {\em projection postulate}, 
because the eigenstate $\phi_{n}$ provoked by the measurement is chosen 
by the projection $E_{n}$ onto the corresponding eigenspace so that
$\phi_{n}=E_{n}\psi/\|E_{n}\psi\|$ for the initial state $\psi$.
It is clear that statements (B1) and (B2) imply (A1) and (A2) as
special cases.  It follows from (B1) and (B2) that the nonselective
measurement of $X$ leads to the state change such as
\begin{equation}
\rho\mapsto\sum_{n}E_{n}\rho E_{n}.
\end{equation}

\section{Interaction between the object and the apparatus}
\label{se:7}

\sloppy
We shall turn to the discussion on the measurement of the discrete
observable $A=\sum_{n}a_{n}|\phi_{n}\rangle\langle\phi_{n}|$ with 
nondegenerate eigenvalues.
In section \ref{se:4}, we have concluded that the state change (1)
in the nonselective measurement must be caused by the interaction
between the object and the apparatus.
Then, what is this interaction?

Let us suppose that the object is in the eigenstate $\phi_{n}$
of the observable $A$ pertaining to the eigenvalue $a_{n}$.
By the statistical formula, the outcome is $a_{n}$ with probability one.
Hence, the measurement changes the position of the pointer in the
apparatus from the original position to the position $a_{n}$ on the scale.

Let $B$ be the observable corresponding to the position of the pointer in
the apparatus, and $\xi$ the original state of the apparatus.
Generally it is only required \cite[p.~439]{vN55}
that the eigenvalues $\{b_{n}\}$ 
of $B$ are in one-to-one correspondence with the eigenvalues
$\{a_{n}\}$ of $A$, but we assume for simplicity that the
observable $B$ has also the same eigenvalues, 
$a_{1},a_{2},\ldots$, as $A$.
In the Hilbert space of the composite system of the object and
the apparatus, the observables $A$ and $B$ are represented by
the self-adjoint operators $A\otimes 1$ and $1\otimes B$ 
respectively.

The state change due to the interaction is represented by 
a unitary operator $U$ on the Hilbert space of the composite 
system:
\begin{equation}
U: \phi_{n}\otimes\xi\mapsto U(\phi_{n}\otimes\xi).
\end{equation}

According to the statistical formula (A1), the state 
$U(\phi_{n}\otimes\xi)$ after the interaction must be 
the eigenstate of $1\otimes B$ pertaining to the eigenvalue
$a_{n}$.
In fact, the position of the pointer takes the value $a_{n}$
with probability one after the interaction, and this implies 
that the state $U(\phi_{n}\otimes\xi)$ is the eigenvector of
$1\otimes B$ pertaining to the eigenvalue $a_{n}$.

On the other hand, according to the repeatability hypothesis (M)
the state $U(\phi_{n}\otimes\xi)$ is the eigenvector of 
$A \otimes 1$ pertaining to the eigenvalue $a_{n}$.
In fact, if the observer were to measure $A$ in this state again,
then the observable $A$ would be measured twice in succession
and hence the outcome would be $a_{n}$ with probability one.
This implies that the state $U(\phi_{n}\otimes\xi)$ is the 
eigenstate of $A\otimes 1$ pertaining to the eigenvalue $a_{n}$.

Suppose here that the eigenvalues of $B$ are also nondegenerate.
Then the state that satisfies the above two conditions is 
represented by the state vector $\phi_n \otimes \xi_n$,
where $\xi_n$ is an arbitrary eigenvector of $B$ with unit length
pertaining to the eigenvalue $a_{n}$.
In order to represent the measurement the unitary $U$ must satisfy 
the following relation 
\begin{equation}\label{eq:1}
U(\phi_n \otimes \xi) = \phi_n \otimes \xi_n.
\end{equation}

If the unitary operator $U$ satisfies the above condition, then
for the arbitrarily given original state
$\psi = \sum_n c_n \phi_n$ we have by linearity 
\begin{equation}\label{eq:2}
U(\psi \otimes \xi) = \sum_n c_n \phi_n \otimes \xi_n.
\end{equation}
Thus the problem is whether equation (\ref{eq:2}) is a sufficient 
condition for the unitary operator $U$ to represent the measuring
interaction or whether, in other words, equation (\ref{eq:2}) implies
(A1) and (A2) even when $\psi$ is a superposition of the eigenstates
$\phi_{n}$.
If equation (\ref{eq:2}) were not a sufficient condition, further interaction
--- though ineffective in the case where the object is initially in
the eigenstate ---
might be necessary for the explanation leading to the state reduction. 

\section{The Orthodox view}
\label{se:8}

The conventional approach adopted by most of 
the text books on measurement theory, the so-called orthodox view, is
negative about the above problem.  The orthodox view holds
that (\ref{eq:2}) does not imply the measurement axiom (A2).  The argument
runs as follows.

The state transformation by the unitary $U$,
\begin{equation}\label{eq:a}
(\sum_{n}c_{n}\phi_{n})\otimes \xi
\mapsto \sum_{n}c_{n}\phi_{n}\otimes\xi_{n},
\end{equation}
makes a one-to-one correspondence between the state of the 
object and the state of the apparatus.
The state of the object is mirrored by the state of the apparatus,
and the problem of a measurement on the object is transformed into
the problem of an observation on the apparatus \cite{Wig63}.
If the observer observes the pointer position of the apparatus to obtain
the outcome of measurement, 
the state in the right-hand side of (\ref{eq:a}) changes as follows:
\begin{eqnarray}\label{eq:3}
\lefteqn{|\sum_n c_n \phi_n \otimes \xi_n\rangle
\langle\sum_n c_n \phi_n \otimes \xi_n|}
\qquad\nonumber\\
&\mapsto&
\sum_n |c_n|^{2} |\phi_n \otimes \xi_n\rangle
\langle \phi_n \otimes \xi_n|. 
\end{eqnarray}
The state change in (\ref{eq:3}) is derived by the projection postulate
(B2) applied to the state change caused by 
the measurement of the observable
$1\otimes B$ of the composite system.
According to the projection postulate (B2), the
new state is the mixture of the states $\phi_n \otimes \xi_n$,
and hence when the outcome $a_{n}$ is obtained, a system in the  
state $\phi_n \otimes \xi_n$ is selected from the ensemble described
by the right-hand side of (\ref{eq:3}):
\begin{equation}\label{eq:c}
\sum_n |c_n|^{2} |\phi_n \otimes \xi_n\rangle\langle \phi_n \otimes \xi_n| 
\mapsto
|\phi_n \otimes \xi_n\rangle\langle \phi_n \otimes \xi_n|.
\end{equation}
Finally, the composite system is in the state $\phi_n \otimes \xi_n$,
and this implies that the object is led to the state $\phi_n$.

Nevertheless, if we describe further the measuring process which 
leads to the state change (\ref{eq:3}) in terms of the coupling with
the second apparatus, having an orthonormal system $\{\xi'_{n}\}$ 
and being prepared in a state $\xi'$, measuring 
the pointer position in the first apparatus, 
then instead of (\ref{eq:3}) we have the following
state change:
\begin{equation}\label{eq:4}
(\sum_n c_n \phi_n \otimes \xi_n)\otimes\xi'
\mapsto\sum_n c_n \phi_n \otimes \xi_n \otimes \xi'_{n}.\quad
\end{equation}
From this state change, we cannot conclude that the measurement leads 
the object with the outcome $a_{n}$ to the state $\phi_{n}$.
The original problem of explaining the state reduction caused by
the first apparatus is not solved but only transferred to the the problem
of explaining the state reduction caused by the second apparatus
\cite[Section 11.2]{Jam74}.
This vicious circle is often called 
{\em von Neumann's chain}.

\section{Inconsistency of the orthodox view}
\label{se:9}

A difficulty in the orthodox view is to apply the projection postulate
to the object-apparatus composite system 
in order to show that the state of the object that leads to the outcome
$a_{n}$  is in the state $\phi_{n}$ at the time
just after measurement.
This argument suffers from the circular argument that assumes the 
rule of state reduction for the composite system in order to explain 
the rule of state reduction for the object.
The conventional studies of measurement theory, however, have not
detected any physical inconsistency or empirical inadequacy
of the above argument in
the orthodox view and have aimed at circumventing
the above circular argument, an epistemological difficulty,
by adding, for instance, the element of macroscopic nature of the 
measuring apparatus \cite{MN80a,Zur81}.

In fact, the above argument leading to the state reduction has been 
generalized to the following argument for any measurements to determine the
state change caused by measurement conditional upon the outcome:
Let us given the initial state $\rho$ of the object, the initial state of the
apparatus $\sigma$, and the unitary evolution operator $U$ of the
object-plus-apparatus.
Then, compute the state of the composite system
just after the interaction as $U(\rho\otimes\sigma)U^{\dagger}$ and
apply the projection postulate to the measurement of the 
pointer observable $B$ in the apparatus, 
and the state $\rho_{a_{n}}$ just after
the measurement conditional upon the outcome $a_{n}$ is given by
\begin{equation}
\rho_{a_{n}}=\frac{
\mbox{\rm Tr}_{{\bf A}}[(I\otimes E^{B}(a_{n}))
U(\rho\otimes\sigma)U^{\dagger}(I\otimes E^{B}(a_{n}))]}
{\mbox{\rm Tr}[(I\otimes E^{B}(a_{n}))U(\rho\otimes\sigma)
U^{\dagger}(I\otimes E^{B}(a_{n}))]},
\end{equation}
where $E^{B}(a_{n})$ is the projection operator onto the eigenspace of
$B$ corresponding to the eigenvalue $a_{n}$ and 
$\mbox{\rm Tr}_{{\bf A}}$ stands for
the partial trace over the Hilbert space of the apparatus.
This unitary-evolution-plus-projection-postulate argument has been
a standard argument for determining the general state reduction,
see for example
\cite{Kra71,kra83,Cav86,BK92,BGL95}.

The purpose of this section is, despite the conventional arguments,
to show that the orthodox view suffers from a serious physical inconsistency
concerning the causality between the reading of the outcome and
the state reduction.

In order to explain the rule of state reduction in terms of 
the time evolution of the object-plus-apparatus,
it is necessary to clarify the meanings of the words the ``time
of measurement'' and the ``time just after measurement'' in the context
as to what happens in the object and the apparatus.
Let us suppose that one measures an observable 
$A=\sum_{n}a_{n}|\phi_{n}\rangle\langle\phi_{n}|$ of the object in the state 
$\psi$ at the time $t$.
The measurement, carried out by an interaction with the apparatus,
takes finite time $\Delta t>0$.
Thus, the object interacts with the apparatus from the time $t$
to $t+\Delta t$ and is free from the apparatus after
the time $t+\Delta t$.
It follows that the time of measurement is the time $t$,
the time just after measurement is $t+\Delta t$, and that
the object is in the state $\psi$ at the time of measurement.
The statistical formula (A1) means, in this case, that the observer
obtains the outcome  $a_{n}$ with probability 
$|\langle\phi_{n}|\psi\rangle|^{2}$.
The measurement axiom  (A2) means that the object that leads to the
outcome $a_{n}$ is in the state $\phi_{n}$ at the time $t+\Delta t$.
Moreover, the repeatability hypothesis (M) means that if the observable
$A$ is measured at the time $t_{1}=t+\Delta t$ again in the 
same object then the outcome coincides with the one obtained by
the measurement of $A$ at the time $t$.

In the orthodox view, the state changes given by (\ref{eq:a}) 
and (\ref{eq:3}) 
represent dynamical changes of the system, and the state change 
(\ref{eq:c}) 
represents a change of the knowledge of the observer.
The state change 
(\ref{eq:a}) represents the interaction between the object and the apparatus.
The state change (\ref{eq:3}) represents the interaction between the 
``apparatus''
and the ``apparatus measuring the apparatus''.
It follows that the state change (\ref{eq:a}) shows that the 
object-plus-apparatus is in the state 
$(\sum_{n}c_{n}\phi_{n})\otimes \xi$ at the time $t$ and in the state 
$\sum_{n}c_{n}\phi_{n}\otimes\xi_{n}$ at the time $t+\Delta t$.

Suppose that the state change (\ref{eq:3}) takes time $\tau$.
Then, it is at the time $t+\Delta t+\tau$ that
the object-plus-apparatus turns to be in the state described by
the right-hand side of (\ref{eq:3}).
Since the state change (\ref{eq:c}) represents the change of 
the observer's knowledge, it does not accompany the change of time
so that at the time $t+\Delta t+\tau$ the object turns
to be in the state $\phi_{n}$ as the result of the state reduction.
In other words, the orthodox view leads to the conclusion that
the state reduction occurs at the time $t+\Delta t+\tau$
which is later in time $\tau$ than the time $t+\Delta t$
just after measurement.
Thus, if $\tau$ is not negligible in the time scale of the time
evolution of the object then this contradicts the measurement
axiom that the state reduction leaves the system in the state $\phi_{n}$
at the time $t+\Delta t$.

Since the object is free from the apparatus after the time 
$t_{1}=t+\Delta t$, one can make the object
interact with the second apparatus at the time $t_{1}$.
If this apparatus also measures $A$, according
to the repeatability hypothesis it is predicted, 
and will be confirmed by experiments, 
that the outcome from the first apparatus and the outcome 
from the second are always the same.  But, this fact cannot be
explained by the orthodox view which concludes that the state 
reduction occurs at the time $t_{1}+\tau$. 

Is $\tau$ negligible in the time scale of the time evolution of 
the object?
In general, the process of the state change (\ref{eq:3}) is regarded as
a process in which a macroscopic instrument operates or a 
directly-sensible variable feature is 
produced --- otherwise, the state change 
in the apparatus measurement might not 
necessarily satisfy the repeatability
hypothesis --- and hence the duration $\tau$ of this process cannot be
negligible in the time scale of the time evolution of the microscopic
measured object.

Consider, for instance, the experiment in which the light 
is scattered by an atom in a low intensity atom beam.
Regarding the paths before the collision as known,
the measurement of the path of the photon after the collision
suffices to determine the point of scattering.
In order to measure the position of the atom (at the time
of collision) twice in succession in this method, 
suppose to use two nearly placed light beams I and II; see
FIG.~\ref{fig:1}.
Suppose that the atom interacts with the beam I from 
the time $t$ to $t'$ and with the beam II from 
$t_{1}$ to $t'_{1}$ and that $t_{1}-t'$ is so small that
the time evolution of the atom in this period can be 
neglected --- hence, we  can put $t_{1}=t'$.
Suppose that the photon scattered from the beam I is
detected by a photoelectric detector at the time $t''$, and
the one from the beam II is detected by another photoelectric detector
at the time $t''_{1}$.
In this experiment, the collision is accomplished in quite
a short time and the photoelectric detectors are necessary 
to place sufficiently far from the light beams, 
so that it is taken for granted in scattering theory 
\cite[p.~375]{Mes59a} that
\begin{equation}\label{eq:6.1}
t'-t\ll t''-t'.
\end{equation}

It is taken for granted from the Compton-Simons experiment that 
there is the uncertainty of the position at which the beam I is 
scattered depending on the initial state of the atom 
but that the position of the scattering from the beam II 
is always near the position of the scattering from the beam I.
It follows that this experiment can be considered as the position 
measurement of the atom
satisfying the repeatability hypothesis \cite[pp.~212--214]{vN55}.
In this example, the state change (\ref{eq:a}) corresponds to the 
interaction between the atom and the light beam,
and hence we have $\Delta t=t'-t$.
On the other hand, the apparatus corresponds to the scattered 
photon, the state change (\ref{eq:3}) corresponds to the process
including 
the free propagation of the photon after scattering and 
the interaction between the photon and the photoelectric detector,
and hence we have $\tau=t''-t'$.
Thus, from (\ref{eq:6.1}) we have
\begin{equation}
\Delta t\ll \tau,
\end{equation}
and consequently we cannot neglect $\tau$.
This means that the orthodox view claims that the state reduction
of the atom occurs after the photon is detected 
despite the fact that the state reduction
of the atom occurs just after the scattering of the
light.

The inconsistency of the orthodox view is in the claim of causality
between the reading of the outcome and the state reduction 
such that the state change  (\ref{eq:3}) of the 
composite system causes the state reduction of the object system.
It is obvious, however, that such causality does not exists,
since the result, the state reduction, occurs before the cause, 
the reading of the outcome or the manifestation of the 
directly-sensible variable feature.
It is not the case that the observer's knowing or reading of the outcome 
at the time $t''=t+\Delta t+\tau$ causes the state 
reduction of the object at the time $t'=t+\Delta t$.
But, it is the case that by knowing or reading of the outcome 
at the time $t''$ the observer obtains the information 
to determine the state of the object at the time $t'$.
The orthodox view confuses the time at which the outcome of measurement
is obtained and the time at which the object is left in the state 
determined by the outcome. 
Or, in other words, it confuses the time just after the reading of the
outcome and the time just after the interaction between the 
object and the apparatus.
There is no causality relation between the outcome and the state
just after measurement but there is coincidence between them
yielded by the measuring interaction.

\section{New interpretation}
\label{se:10}

Our new interpretation presented in the following does not
includes the inconsistency of the orthodox view discussed
in the previous section.
Moreover, the state reduction can be explained only from (\ref{eq:a}) 
without assuming the process (\ref{eq:3}) so that
the circular argument of the von Neumann chain is circumvented.

As in the preceding section, 
suppose that the observer measures the observable
 $A=\sum_{n}a_{n}|\phi_{n}\rangle\langle\phi_{n}|$ of the object in the
state $\psi$ at the time $t$.
The object interacts with the apparatus from the time
$t$ to $t+\Delta t$ and is free from the apparatus
after the time $t+\Delta t$.
The repeatability hypothesis (M) means that if the observer measures
$A$ at the time  $t_{1}=t+\Delta t$ again then the outcome
coincides with the outcome of the measurement at the time
$t$.
As shown previously, the measurement axiom (A2) is equivalent to
the repeatability hypothesis (M).
Hence, if the statistical formula (A1) and
the repeatability hypothesis (M) is derived from (\ref{eq:2}), 
it is demonstrated that the state
reduction is derived from (\ref{eq:2}).
Let $\Pr\{B(t+\Delta t)=a_{n}\}$ be the probability of
obtaining the outcome $a_{n}$ when the pointer position 
$B=\sum_{n}a_{n}|\xi_{n}\rangle\langle\xi_{n}|$ is measured at the time
$t+\Delta t$.
By (\ref{eq:2}) and the statistical formula (B1) for the degenerate 
observable
$1\otimes B$ we have
\begin{eqnarray*}
\lefteqn{\Pr\{B(t+\Delta t)=a_{n}\}}\\
&=&\langle \sum_k c_k \phi_k \otimes \xi_k|
(1\otimes|\xi_{n}\rangle\langle \xi_{n}|)
|\sum_{k} c_{k} \phi_{k} \otimes \xi_{k}\rangle\\
&=&|c_{n}|^{2}.
\end{eqnarray*}
Let $\Pr\{A(t)=a_{n}\}$ be the probability that the measurement of 
$A$ at the time $t$ leads to the outcome $a_{n}$.
Since this outcome is obtained as the outcome of the measurement of
$B$ at the time  $t+\Delta t$, we have
\begin{equation}
\Pr\{A(t)=a_{n}\}=\Pr\{B(t+\Delta t)=a_{n}\}=|c_{n}|^{2}.
\end{equation}
Thus if we regard this process as the measurement of $A$ --- namely,
if we interpret the outcome of the measurement of $B$ at the
time $t+\Delta t$ as the outcome of the measurement
of $A$ at the time $t$ --- then 
it is shown that this measurement satisfies the the statistical 
formula (A1).

We shall show that this measurement satisfies the measurement
axiom (A2) in the following.
Since the measurement axiom (A2) is equivalent to the repeatability
hypothesis (M), we need only to show that this measurement satisfies
the repeatability hypothesis (M).

In order to show the last statement, it suffices to show that if the 
observer measures $A$ again at the time
$t_{1}=t+\Delta t$, immediately after the first measurement, 
then the outcome of the first measurement at the time  
$t$ and that of the second at the time $t_{1}$ are always 
the same.
Let $\Pr\{A(t)=a_{n},A(t_{1})=a_{m}\}$ be the joint probability 
that the first outcome is $a_{n}$ and the second outcome
is $a_{m}$.

The outcome of the first measurement of $A$ at the time $t$ is
the same as the  outcome of the measurement of the pointer position
$B$ at the time $t_{1}=t+\Delta t$.
Since the measurements of $B$ and $A$ at the time $t_{1}$ does not
interferes each other, the joint probability distribution
of the outcomes of these two measurements satisfies the statistical
formula for the simultaneous measurements:
\begin{eqnarray*}
\lefteqn{\Pr\{B(t_{1})=a_{n},A(t_{1})=a_{m}\}}\qquad\\
&=&|\langle\phi_{m}\otimes\xi_{n}|
\sum_{k}c_{k}\phi_{k}\otimes\xi_{k}\rangle|^{2}\\
&=&\delta_{m,n}|c_{n}|^{2}.
\end{eqnarray*}
Thus if $m\not=n$ then we have
\begin{eqnarray*}
\lefteqn{\Pr\{A(t)=a_{n},A(t_{1})=a_{m}\}}\qquad\\
&=&\Pr\{B(t_{1})=a_{n},A(t_{1})=a_{m}\}\\
&=&0.
\end{eqnarray*}
It follows that the outcome of the first measurement
and that of the second are always the same.
Therefore, this process satisfies the repeatability hypothesis (M)
and hence satisfies the measurement axiom (A2).

We have thus demonstrated that the unitary operator $U$ satisfying
(\ref{eq:2}) represents the interaction between the object and the
apparatus and leads to the state reduction in the object
at the time just after measurement.

\section{Concluding remarks}
\label{se:11}

It follows from the basic postulate requiring the state reduction
that even in the case where the observer obtains no information
from the measurement, namely the case of nonselective measurement,
the state of the system changes.
This change is not accompanied with the change of knowledge so that
it is purely dynamical, but it is irreversible so that it cannot be
described by the Schr\"{o}dinger equation of the object.
The only way to explain this by the rules of quantum mechanics is
to derive this change from the Schr\"{o}dinger equation of a 
larger system than the object, which describes the interaction between
the object and the apparatus. 
This interaction is turned on during a finite time interval,
from the time  $t$ of measurement to the time $t+\Delta t$ just
after measurement.  After the object-apparatus interaction,
the object turns
to be free from the apparatus again.
Thus, the state reduction describes the state change from
the time $t$ to the time $t+\Delta t$ conditional upon the outcome of
measurement.

The state change in the nonselective
measurement is derived without any difficulties from
the interaction between the object and the apparatus.
In fact, the state change (\ref{eq:0}) is explained 
as the open system dynamics
of the object yielded by the unitary evolution of the object-apparatus
composite system described by the unitary $U$ in (\ref{eq:2}), i.e., 
\begin{equation}
\sum_{n} |\langle\phi_n|\psi\rangle|^2\,|\phi_n\rangle\langle\phi_n|
=\mbox{\rm Tr}_{{\bf A}}[U|\psi\otimes\xi\rangle\langle\psi\otimes\xi|
U^{\dagger}],
\end{equation}
where $\mbox{\rm Tr}_{{\bf A}}$ is the partial trace over the Hilbert 
space of the
apparatus \cite[p.~136]{Dav76}. 
The problem is to explain the change of state dependent on the
outcome, namely the state reduction.
The answer of the orthodox view to this problem is that
the state reduction of the object is resulted from 
the state reduction of the object-plus-apparatus caused by
the measurement of the pointer position carried out
after the object-apparatus interaction.
Applying the above argument that state reduction needs the time for
the nonselective measurement to the measurement of the pointer 
position, it is concluded that the state reduction, explained
by the orthodox view, occurs apparently later than the time $t+\Delta t$
at which the state reduction should occur.
This time difference leads to the detectable difference
as to whether the outcomes obtained by measuring the same object
twice in succession satisfy the repeatability hypothesis,
and hence we can conclude that the inconsistency of the  
the orthodox view can be tested by an experiment.

The photon scattering experiment from the atom beam 
in an atom interferometer has been realized already by 
Chapman {\em et al.\/} \cite{CHLSRSP95}.
The double scattering {\em gedanken} experiment suggested in Section
\ref{se:9} will be realized in future along with a similar 
experimental setting with the additional second laser beam for
the repeated measurement of the point of scattering of a single
atom, if a conceivable difficulty can be circumvented 
in distinguishing the case where two detected photons from the two 
beams have been scattered by a single atom from the other cases.

\begin{figure}
\caption{Repeated position measurements by photon scattering from an atom.}
\label{fig:1}
\end{figure}

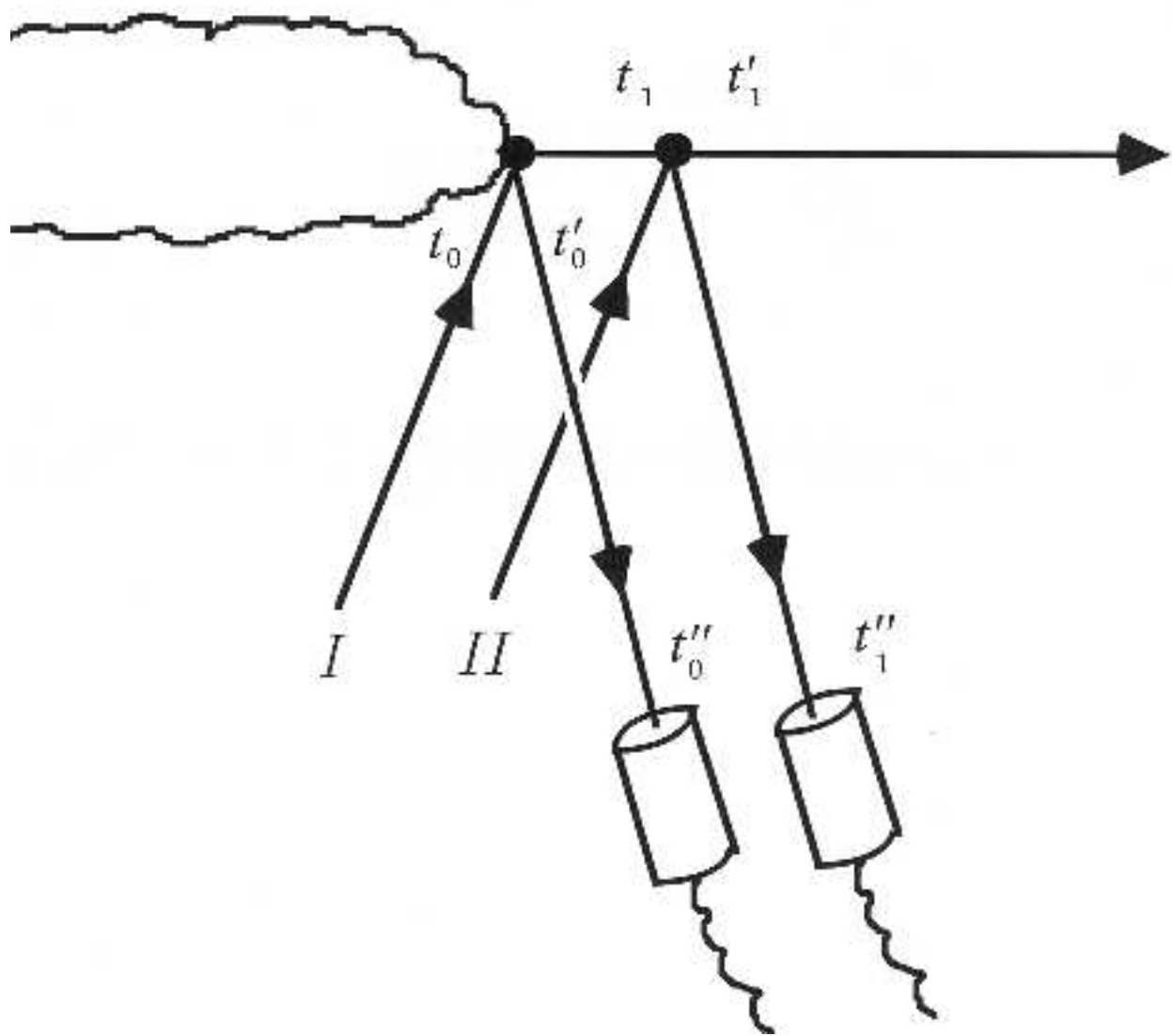
\end{document}